\documentclass{PoS}

\def\one{{\bf 1}}

\def\five{{\bf 5}}
\def\fiveb{{\overline{\bf 5}}}
\def\ten{{\bf 10}}
\def\tenb{{\overline{\bf 10}}}

\title{``Invisible'' axion rolling through the QCD phase transition}

\ShortTitle{``Invisible'' Axion \& QCD}

\author{\speaker{Jihn E. Kim}\thanks{ This work is supported in part by the National Research Foundation (NRF) grant  NRF-2018R1A2A3074631 and by the IBS (IBS-R017-D1-2014-a00).}\\
        Department of Physics, Kyung Hee University, 26 Gyungheedaero, Dongdaemun-Gu, Seoul 02447, Republic of Korea, and\\
Center for Axion and Precision Physics Research (Institute of Basic Science), KAIST Munji Campus, 193 Munjiro,  Daejeon 34051, Republic of Korea \\
        E-mail: \email{jihnekim@gmail.com}}

\author{ Se-Jin  Kim\thanks{This work is supported in part by the National Research Foundation (NRF) grant  NRF-2015aR1D1A1A09059301.}\\
        Department of Physics, Kyung Hee University, 26 Gyungheedaero, Dongdaemun-Gu, Seoul 02447,   Republic of Korea }

\abstract{The origin of ``invisible'' axion in four dimensional effective beyond-standard models from string compactification is discussed and   its  refined passover through the QCD phase transition is presented toward a reliable estimate of the current axion energy density in terms of the initial misalignment angle $\bar{\theta}_1$. The explicit examples are presented in a flipped SU(5) GUT model. This allows to introduce a flavor symmetry through string compactification, and hence we also comment on the source of flavor symmetries from string compactification and attempts to fit  the resulting Yukawa couplings  to the observed Cabibbo-Kobayashi-Maskawa  and  Pontecorvo-Maki-Nakagawa-Sakata matrices.}

\FullConference{Corfu Summer Institute 2018 "School and Workshops on Elementary Particle Physics and Gravity"\\
		(CORFU2018)\\
		31 August - 28 September, 2018\\
		Corfu, Greece}

\begin{document}

%%%%%%%%%%
\section{Introduction}

The Planck scale $M_{P}\simeq 2.4\times 10^{18\,}$GeV seems to be the definition of physics scale. So, when new phenomenon at low energy seemed to be present, first  plausible attempts might be to understand them in terms of known physics of that time. When neutral currents were announced, I tried it along this road \cite{Mathur74,KimMM76,KimMM78}.  It was noted that a magnetic moment proportional to the neutrino mass can be obtained even without the assumption of two neutrino helicities long time ago \cite{KimMM76}. This is because the chiral structures of fermion mass and magnetic moment are the same, except for the  electromagnetic coupling $e$. Chiral structure is an important information of low energy fermions. Recently, when no new physics is announced at the TeV scale, I followed this line of argument on chiral structure to see a possibility of light particles which are hidden from the LHC \cite{Kim17Ch}.

The first chiral structure proposed at low energy is the ``V-A'' theory of charged current (CC) weak interactions. Independent left(L)- and right(R)-handed fermion representations with gauge principle only for quarks and leptons can introduce, by accident, extra global symmetries such as the baryon and lepton numbers in the standard model (SM). If global symmetries are introduced, their destiny is to be broken at some scale \cite{Kim18IJMPA}. The well-known global symmetry is the Peccei-Quinn (PQ) global symmetry \cite{PQ77} which is broken dominantly  by the QCD anomaly. The PQ symmetry could have been broken by the fields at the electroweak scale only \cite{Weinberg77} but there was no hint that such breaking takes place at the electroweak scale by the time 1978 \cite{Peccei78}. So, it was necessary to introduce weak interaction singlet(s) beyond the SM (BSM). A phase among these singlets \cite{Kim79} is the so-called ``very light'' \cite{Kim79} or ``invisible'' axions which are named as the KSVZ axions \cite{Kim87prp,SVZ80} and DFSZ axioins \cite{DFSZ}. The BSM singlets must be introduced above the TeV scale \cite{Kim79}. Here again, the chiral symmetry is crucial to understand how such singlets survive down below the Planck scale $M_P$.
  
%%%%%%%%%%
\section{U(1)$_{\rm anom}$: the source of ``invisible'' axion}
Here we glimpse a scenario how the PQ symmetry breaking scale survives down  to the intermediate scale $M_I$ from string compactification. Along this line, we assign the PQ quantum numbers  to all the fields, including BSM singlets. The relevant U(1) symmetry is U(1)$_{\rm anom}$ whose charges will be denoted as $Q_{\rm anom}$. Since the SM fields have respective  $Q_{\rm anom}$ quantum numbers, it can act as the quantum number of a  flavor symmetry \cite{KimKyaeNam17}. So, this idea relates the ``invisible'' axion with the flavor symmetry in the SM. Related to this, we will briefly comment on flavors in Sec. 6.

The mechanism behind lowering the PQ symmetry breaking scale is the so-called 't Hooft mechanism \cite{Hooft71,KimPatras17}: 
\begin{itemize}
\item[]``If a global symmetry and  a gauge symmetry are broken by the vacuum expectation value (VEV) of one complex scalar field, then the gauge symmetry is broken but a global symmetry remains unbroken''.
\end{itemize}
This is obvious because the gauge boson must obtain mass by the Brout-Englert-Higgs-Guralnik-Hagen-Kibble mechanism and one continuous shift symmetry from the original two angle directions cannot be broken. This unbroken shift symmetry is a global symmetry because there does not exist the  corresponding gauge boson below the VEV. This fact was noted long time ago
 \cite{Kim88} in string compactification.
 
 In spontaneously broken gauge models, the signal for the gauge boson mass arises from the mixing term of the longitudinal mode $a$ (the phase of the complex scalar field $\phi$) and the gauge field $A_\mu$
 \begin{equation}
 |D_\mu \phi|^2=\frac{1}{2}(\partial_\mu a)^2-g Q_a A_\mu\partial^\mu a+\frac{g^2}{2}Q_a^2 v^2 A_\mu^2
 =\frac{g^2}{2}Q_a^2 v^2(A_\mu- \frac{1}{gQ_a v}\partial_\mu a)^2 
\end{equation}
and the longitudinal degree disappears by redefining the longitudinal component of $A_\mu$ as  $A_\mu'=A_\mu-  \frac{1}{gQ_a v}\partial_\mu a$.
 
In compactifications of the heterotic string 10D$\to$4D, the model independent (MI) axion component comes from the tangential component of $B_{MN}(M,N=1,2,\cdots,10)$  \cite{GS84,Witten84}: $B_{\mu \nu}(\mu,\nu=1,2,3,4)$,
 \begin{equation}
 H_{\mu\nu\rho}= M_{MI}\epsilon_{\mu\nu\rho\sigma}\partial^\sigma a_{MI}.
\end{equation}
The rank 16 gauge group E$_8\times$E$_8'$ can produce many U(1)'s beyond the SM gauge group. If all U(1) do not have any gauge anomaly, then the shift symmetry $a_{MI}\to a_{MI}+$(constant) is broken at the compactification scale. On the other hand, if there appears an anomalous U(1) from E$_8\times$E$_8'$ \cite{AnomU187},  U(1)$_{\rm anom}$, then the gauge boson $A_\mu^{\rm anom}$ corresponding to U(1)$_{\rm anom}$ obtains mass.  Here,  $H_{\mu\nu\rho}$ couples to the anomalous gauge boson $M_{MI}A_\mu^{\rm anom} \partial^\mu a_{MI}$ and the 't Hooft mechanism works as shown in   \cite{KimKyaeNam17}. So, in the string compactification models with a 4D anomalous U(1) gauge symmetry, the original shift symmetry of the MI axion, $a_{MI}\to a_{MI}+$(constant), survives as a global PQ symmetry below the compactification scale. The ``invisible'' axion realized around $10^{11\,}$GeV scale can have this origin of the PQ symmetry from string compactification.
    
%%%%%%%%%%
\section{QCD phase transition}
The QCD phase transition occurs around the critical temperature $T_c\approx 150-165\,$MeV as shown in the lattice gauge theory \cite{ICTP16,Borsanyi16}. Also, we should take into account the evolution of the Universe since our current hadronic phase occured in this Universe, not in a nuclear theory laboratory in a finite volume. The Universe cools down to $T_c$ from a high temperature.  It is in the quark and gluon phase above $T_c$  and  in the hadronic phase below $T_c$. The spin degrees of freedom $g_*$ used in the cosmic evolution are very different above and below $T_c$, and correspondingly the energy and entropy densities also
\begin{eqnarray}
&&\textrm{Before phase transition: }   \rho=\frac{\pi^2}{30}g^i_*T^4=\frac{51.25\pi^2}{30}T^4,~s=\frac{2\pi^2}{45}g^i_*T^3
\label{eq:AboveTc}\\[0.5em]
&&\textrm{After phase transition: } \rho=\frac{\pi^2}{30}g^f_*T^4=\frac{17.25\pi^2}{30}T^4,~s=\frac{2\pi^2}{45}g^f_*T^3\label{eq:BelowTc}
\end{eqnarray}
If we count $g_*$ only for strongly interacting particles, quarks, gluons, and pions, we have  $g^i_*=37$ and $g^f_*=3$. There is a huge gap in the spin degrees around $T_c$ and hence there seems to be no convincing numbers on the QCD phase transition so far. To calculate the current axion energy density reliably, we have to clarify this QCD phase transition problem. Early papers on the axion enrgy density of 1983 \cite{Preskill83} could not consider this problem without the knowledge on the strong interaction at that time, and  hence it is different from the current estimates by a few orders.

The first serious attempt to tackle this QCD phase transition was performed in the mid-1980 in the MIT bag model \cite{DeGrand84}.  In one figure of their study shown in Fig. 1\,(a), evolutions of heat $Q$ and enthalpy $H$ are shown in the pressure vs. temperature plane. These trajectories are different from ours as we see below. Tunneling with an effective potential  $V_{\rm eff}$,  shown in Fig. 1\,(b), was discussed in \cite{KolbBk}.

Recently, the lattice calculation \cite{ICTP16,Borsanyi16} of $T_c$ on the QCD phase transition has attracted a great deal of attention.
Related to the temperature, the discrete time in the Euclidian space corresponds to temperature. When the lattice gauge theory is used above $T_c$, it makes sense because the theory has the gauge symmetry SU(3)$_c$.  The susceptibility calculation jumps sharply at $T_c$ and hence the critical temperature calculated in the lattice computation is used here. There are two issues on the QCD phase transition here. We argue from physical grounds that the lattice computation cannot be extended below $T_c$. The phase below $T_c$ is in the hadronic phase, and using constituent quark masses below $T_c$ on the lattice does not make sense because there is no gauge symmetry.  Below $T_c$, we must use only three (in case of one family with negligible baryon number) pion degrees for hadrons. The phase transition must be calculated in the evolving Universe because our Universe has expanded during the QCD phase transition. In Ref. \cite{DeGrand84}, the evolution effect has been considered also. The phase transition deals with formation of hadronic bubbles and their expansion. Here, we consider the QCD phase transition from the first principles. Our study will show that the phase transition is not the first order but mimics the cross-over phase transitioin. 
%%%%%%%%%%%%%%%%%%%
\begin{figure}[t!]
 {\includegraphics[width=1
 \textwidth]{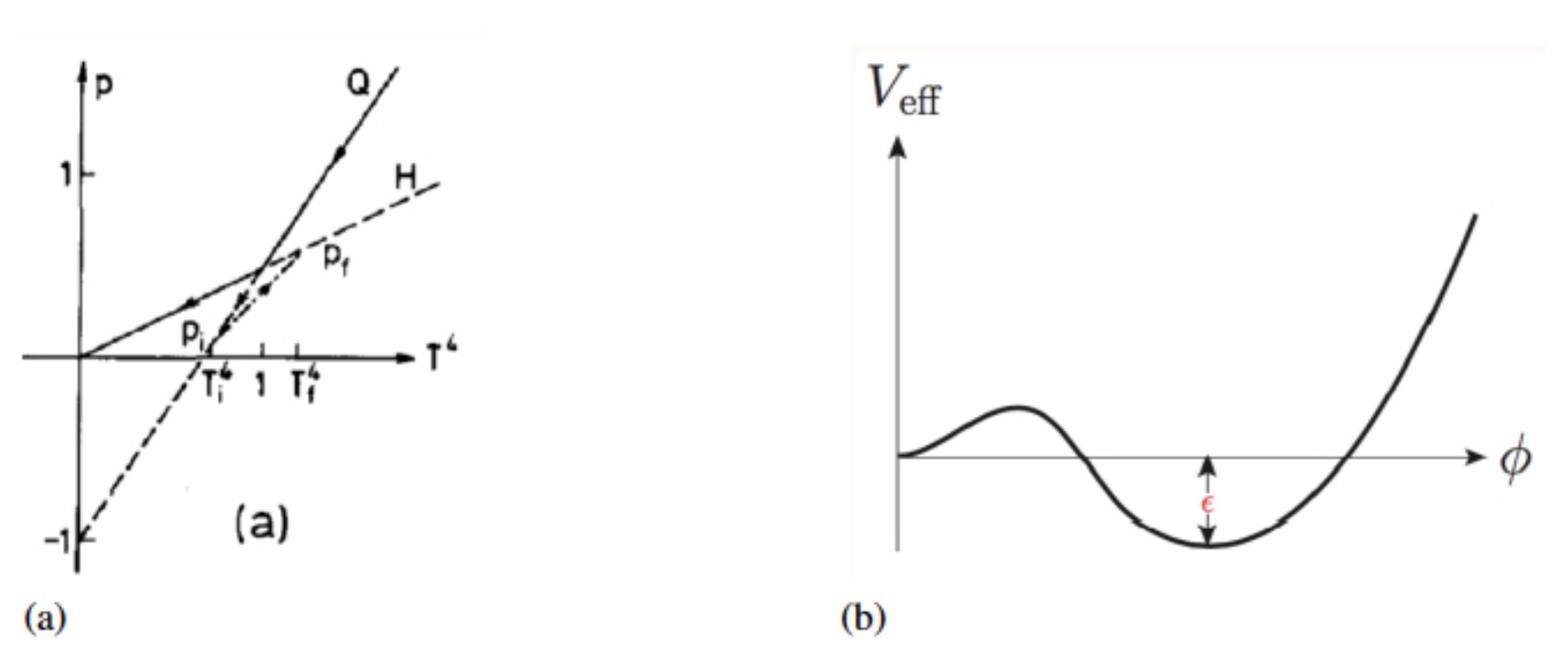}}   
\caption{Early considerations of  QCD phase transitions. (a)  $Q$ and $H$ of Ref. \cite{DeGrand84} and (b) $V_{\rm eff}$ of \cite{KolbBk}.}\label{Fig:QCDphOld}
 \end{figure}
 %%%%%%%%%%%%%%%
  
Firstly, when bubbles form, we consider that typical hadronic size bubbles form initially at rest. This amounts to requiring that pressures $P_{\rm qg}$ in the quark-gluon phase and $P_{\rm h}$ in the hadronic phase are the same at  $T_c$. For the thermodynamic energy variables, therefore, it is better to use the Gibbs free energy $G$ because two independent variables of Gibbs free energy $G$ are $P$ and $T$ whose differential is given by 
 \begin{equation}
dG= -SdT+VdP.
\end{equation}
In statistical mechanics, the $G$ conservation is used for the first order phase transition \cite{HuangBk}. The first order phase transition assumes the true and false vacua as depicted in Fig. 1\,(b). But, we are  not assuming the first order phase transition and just adopt the  conservation of Gibbs free energy. Where is the difference? It is on the calculation of pressure. Above $T_c$, pressure is calculated with the knowledge of spin degrees of freedom of Eq. (\ref{eq:AboveTc}), which is reliable because the QCD coupling constant is small and the approximation of quarks and gluons as point-like blocks is reliable. We know
$P$ of massless quarks and gluons at temperatures $T$, just $\frac13$ of energy
density. Belowe $T_c$, pressure cannot be calculated from  Eq. (\ref{eq:BelowTc})  because the pion waves are of the hadronic scale and the adjacent waves overlap each other.
%%%%%%%%%%%%%%%%%%%
\begin{figure}[t!]
 {\includegraphics[width=0.6
 \textwidth]{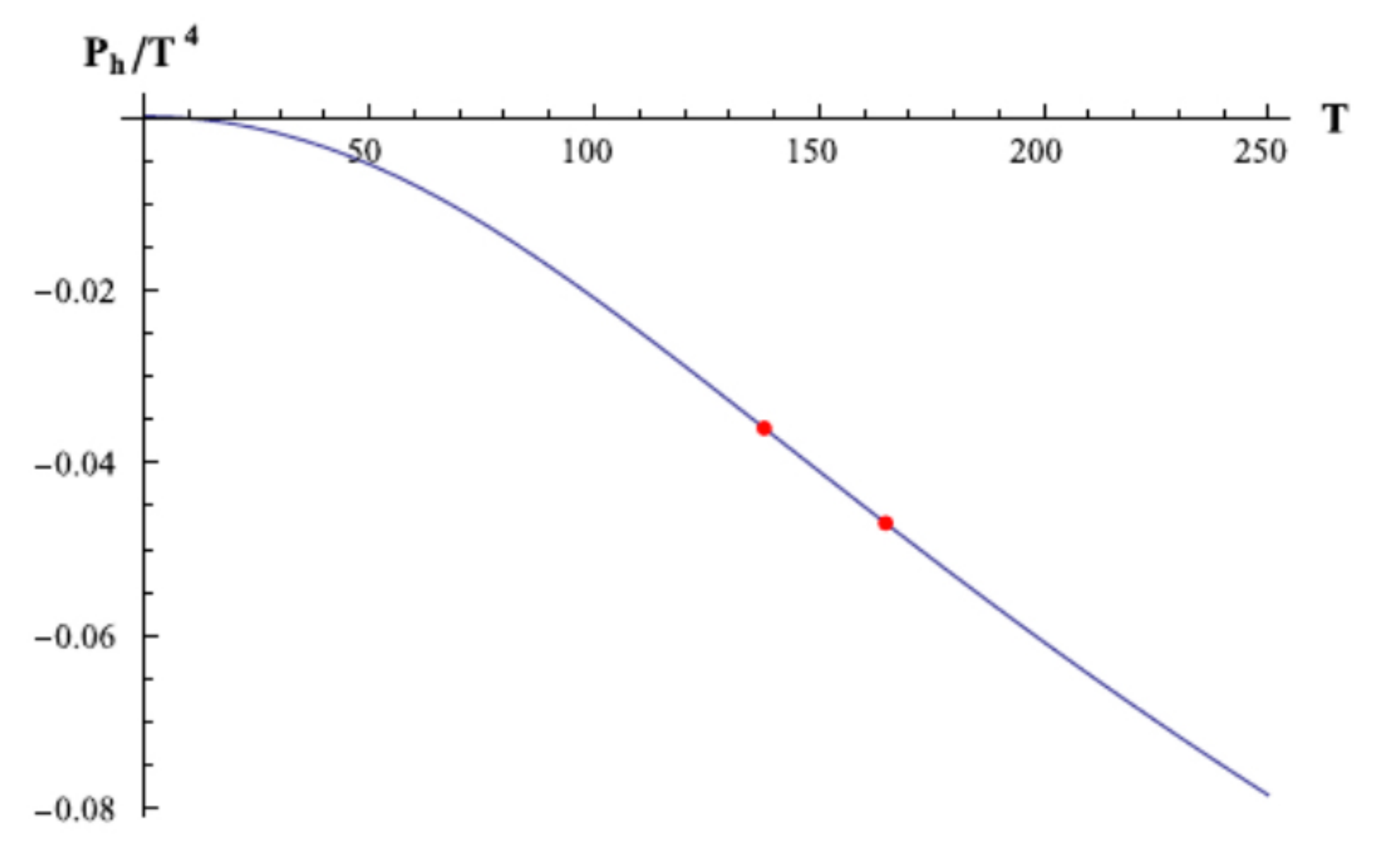}}   \\
\caption{Pressure in the hadronic phase in the $P_h-T$ plane. The right-hand side red dot corresponds to $T_c$.}\label{Fig:Ph}
 \end{figure}
 %%%%%%%%%%%%%%%%%%%%%%%%%%%%%%%%%%
\begin{figure}[b!]
 {\includegraphics[width=0.8
 \textwidth]{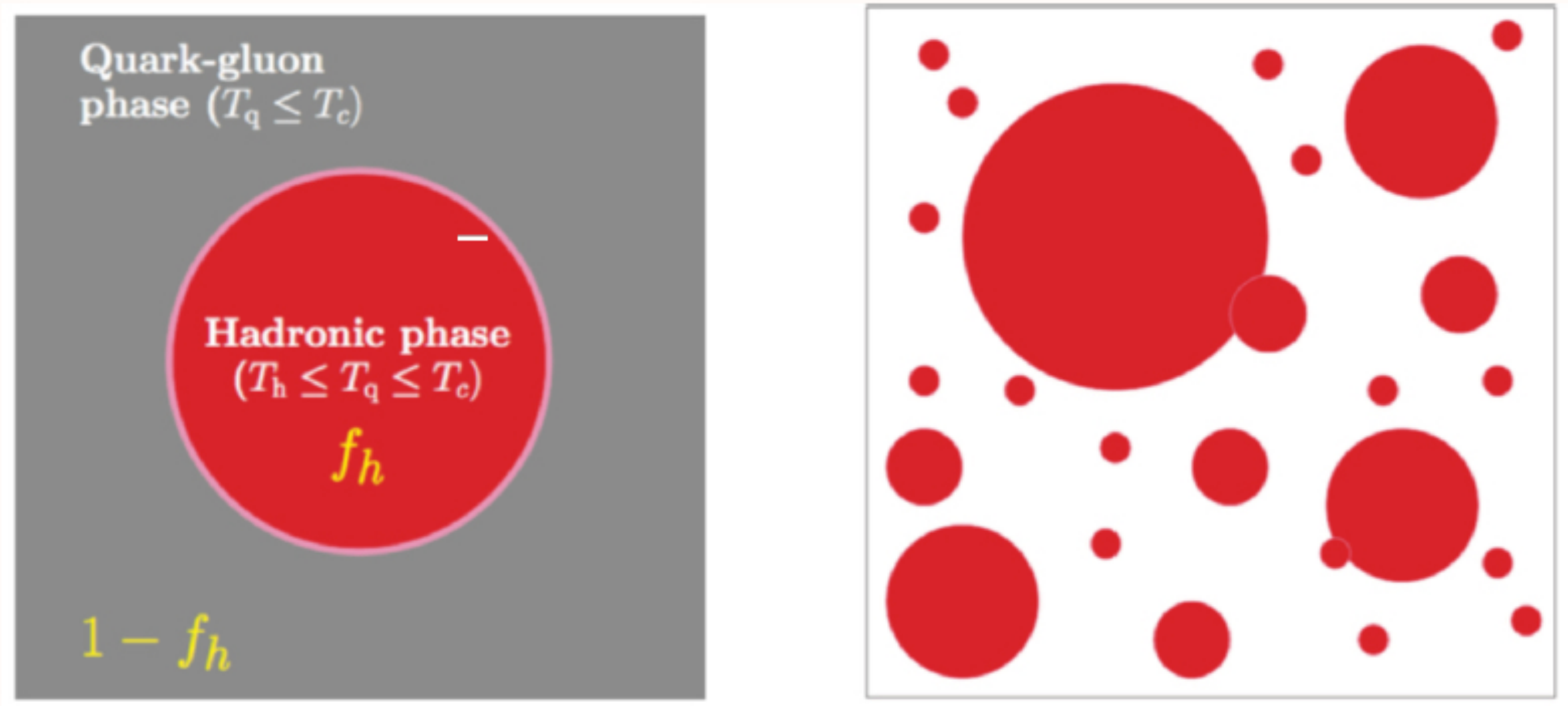}}   \\
\caption{Hadronic bubble(s) formation (in red) in the quark gluon phase background. Left panel:   Right panel: }\label{Fig:QCDbubbles}
 \end{figure}
 %%%%%%%%%%%%%%%

  With the overlapping pion waves, we must calculate pressure $P$ which has been done with the relativistic dispersion relation  \cite{Kim18QCD}, which is shown in Fig. \ref{Fig:Ph}. At $T_c$, a bubble (red) of the pion size is formed in the quark-gluon phase background (gray) as depicted in Fig. \ref{Fig:QCDbubbles}\,(a).  There is no barrier (due to the requirement of the same $P$) to cross over in our case and hence it cannot be the first order phase transition. It is simply the quantum mechanical phenomenon that the pion wave created by the phase transition suddenly extends over to the horizon of the QCD scale.

Second, sometime later these bubbles expand with light velocity in the expanding Universe. Let us introduce a two parameter differential equation for the hadronic fraction $f_h$ as the cosmic time $t$ increases. We must take into account the Hubble expansion in the radiation dominated Universe. We adopt the boundary condition that in the initial formation time $t=0$ there is no overlapping bubbles and also in the final state of the phase transition there is no overlapping because all bubbles are already taken into account.
Our  differential equation for $f_h$ in terms of two parameters ($\alpha$ and $C$) is
\begin{equation}
\frac{df_h}{dt}=\alpha (1-f_h)+\frac{3}{1+Cf_h (1-f_h)(t+R_i)} f_h\label{eq:BubbleEv}
\end{equation}
where $\alpha$ is the conversion rate to hadronic bubbles in the quark-gluon background and the second term takes into account the Hubble expansion of the already formed hadronic bubbles. $R_i$ is the typical size of initially formed hadronic bubbles. This can be taken as the pion size. The conversion rate $\alpha$ in Eq. (\ref{eq:BubbleEv}) is calculated by the Gibbs free energy conservation,
\begin{equation}
(-SdT-PdV)_{\rm quark-gluon~phase}+(-SdT-PdV)_{\rm hadronic~phase}=0
\end{equation}
where we can use $dV_{\rm qg}=-dV_{\rm h}$, or
\begin{equation}
\frac{dV_{\rm h}}{Vdt}  =\frac{S_{\rm qg}-S_{\rm h}}{P_{\rm h}-P_{\rm qg}}  \,\frac{dT}{dt},
\end{equation}
leading to 
\begin{equation}
\alpha= \frac{S_{\rm qg}-S_{\rm h}}{P_{\rm h}-P_{\rm qg}}  \,\frac{dT}{dt}= \frac{-37\pi^2}{45(P_{\rm h}-P_{\rm qg})}  \,\frac{T^6}{\rm MeV},~\textrm{with }T^2t_{\rm [s]}=\rm MeV.
\end{equation}

Figure \ref{Fig:QCDbubbles}\,(b) shows the different scale bubbles some time after the beginning of the QCD phase transition. Overlapping situation is depicted for two bubbles. The situation from the intitial time $t_i$ and the beginning of expansion with light velocity is not analyzed in detail. We just took the beginning time of the light velocity expansion is right after $t_i$.

The QCD scale in axion physics adopts 1\,GeV as a typical scale for the beginning of the misalignment angle. But, we will change this number in this talk due to our better knowledge on the QCD phase transition. Let us first note the relevant expressions of axion mass  in the quark-gluon and hadronic phases,
\begin{eqnarray}
&&\textrm{Quark-gluon phase with }\Lambda_{\rm QCD}:   \approx \frac{ m_u ^2\Lambda^2_{\rm QCD} }{2Z\cos\bar{\theta}+1+Z^2}\,  \sin^2\bar{\theta} , \label{eq:QgVar}\\[0.5em]
&&\textrm{Hadronic phase with }  f_\pi^0~{\rm and}~m_\pi^0:     (f_\pi^0 m_\pi^0)^2\frac{ \sqrt{2Z\cos\bar{\theta}+1+Z^2} }{1+Z}, \label{eq:HadronicVar}
\end{eqnarray}
where $\bar{\theta}=a/f_a$. In Eq. (\ref{eq:QgVar}), we used the Baluni form \cite{Baluni79}, Eq. (5.47) for $\bar{\theta}\simeq 0,\pi$ of \cite{Kim87prp} for vertices of quark loops in the quark-gluon phase, parametrizing the result of the loop integral as the hadronic coupling $  \Lambda^2_{\rm QCD}$. Here, the expression is in terms of parameters  determined by single particle effects. In our case,   we use   Eq. (\ref{eq:QgVar}) to obtain $T_1$ for  $\bar{\theta}\simeq 0,\pi$.  $\bar{\theta}$ near $\pi$ is the relevant region to study a large anharmonic term. And Eq. (\ref{eq:HadronicVar}) is the well-known form in the hadronic phase, where the hadronic parameters in the broken phase of many particles, $f_\pi^0$ and $ m_\pi^0$, are used.
We split the temperature region by four bands as shown below.

%%%%%%
\subsection{Obvious quark gluon phase}

Above the $\rho$ meson mass scale, the phase is obviously in the quark-gluon phase. The  $\rho$ meson mass is roughly twice of the current quark mass and hence it can be considered to be a bound state of a quark and an anti-quark. So, above this scale, spontaneous breaking of chiral symmetry need not be considered and quarks can be relevant degrees. In this quark-gluon phase, the temperature dependence of axion mass has the power $T^{-8.16}$ \cite{Pisarski81}.  Below $T_c$, there is no temperature dependence of axion mass. So, between 1 GeV and $T_c$, we change the temperature power to an interim value $T^{-4.2}$. We choose this cusp point at the scale of $m_\rho$.

%%%%%%
\subsection{Interim phase}

Between the $\rho$ meson mass scale and $T_c$, there is no particle which can be considered as a bound state of a quark  and an anti-quark. K mesons exist but they are considered to be the remnant of chiral symmetry breaking. Furthermore, we do not consider the second family in this talk. Below $T_c$, there is no temperature dependence of axion mass. In this interim region, we use the temperature power  $T^{-4.2}$.

%%%%%%
\subsection{Around $T_c$}

Starting from $T_c$, temperature drops during the QCD phase transition. There are some hadronic phase bubbles as shown in Fig.  \ref{Fig:QCDbubbles} and here we apply the bubble evolution equation (\ref{eq:BubbleEv}). During this phase, the background quark gluon phase can be super-cooled but we keep it at $T_c$. Using the supercooled quark-gluon phase does not change the result very much.

%%%%%%%%%%%%%%%%%%%
\begin{figure}[t!]
 {\includegraphics[width=0.8
 \textwidth]{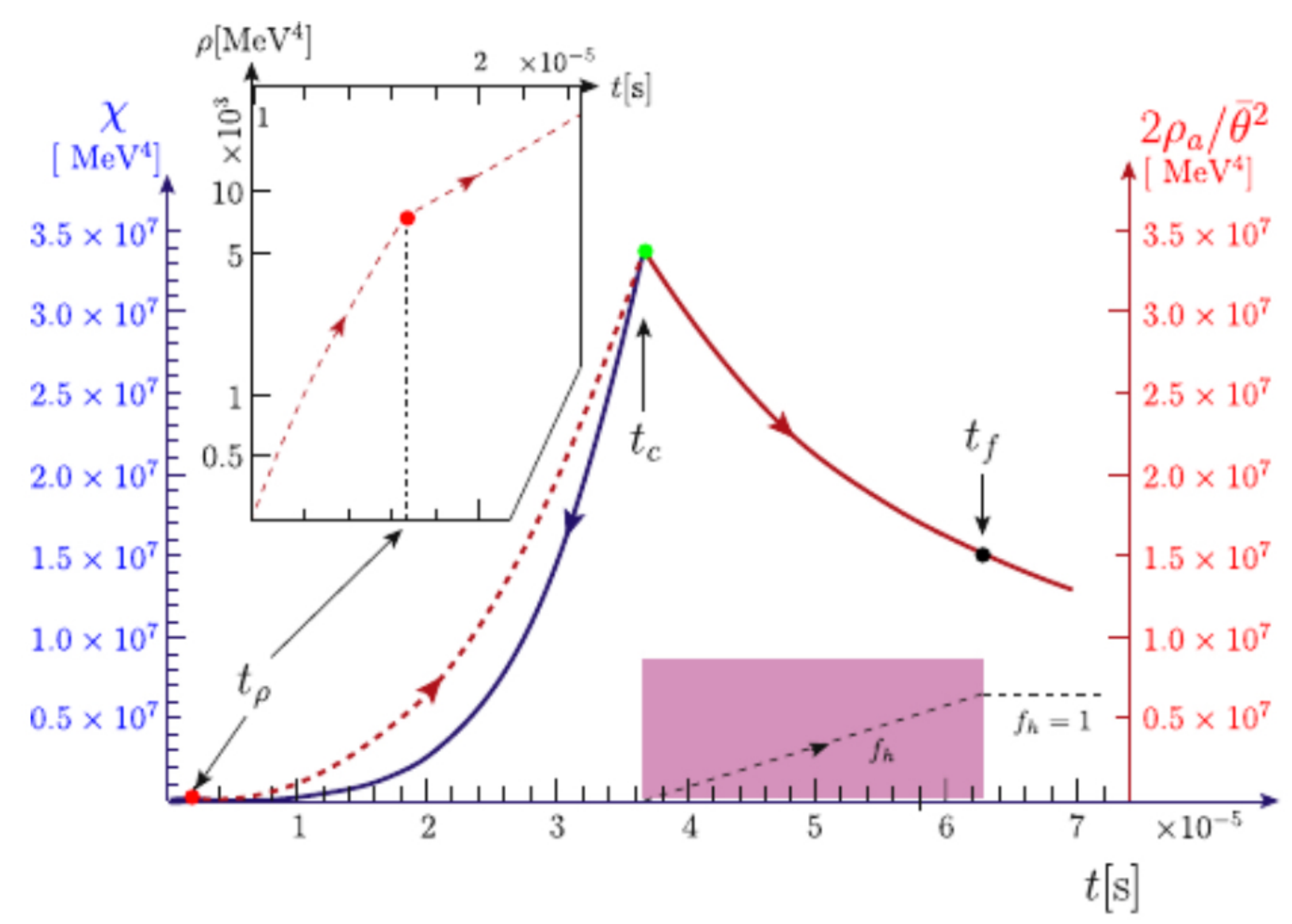}}   \\
\caption{Evolution of susceptibility and axion energy density during the QCD phase transition. }\label{Fig:Overall}
 \end{figure}
 %%%%%%%%%%%%%%%

%%%%%%
\subsection{After the QCD phase transition}

After changing all quark-gluon phase to the hadronic phase, {\it i.e.} obtaining $f_h=1$, the QCD phase transition is finished at time $t_f$. After $t_f$, we solve the evolution of $\bar{\theta}$ \cite{Kim18ThNow}.
 
%%%%%%%%%%
\section{Summary of $\bar{\theta}$ evolution}
%%%%%%%%%%%%%%%%%%%
\begin{figure}[t!]
 {\includegraphics[width=0.6\textwidth]{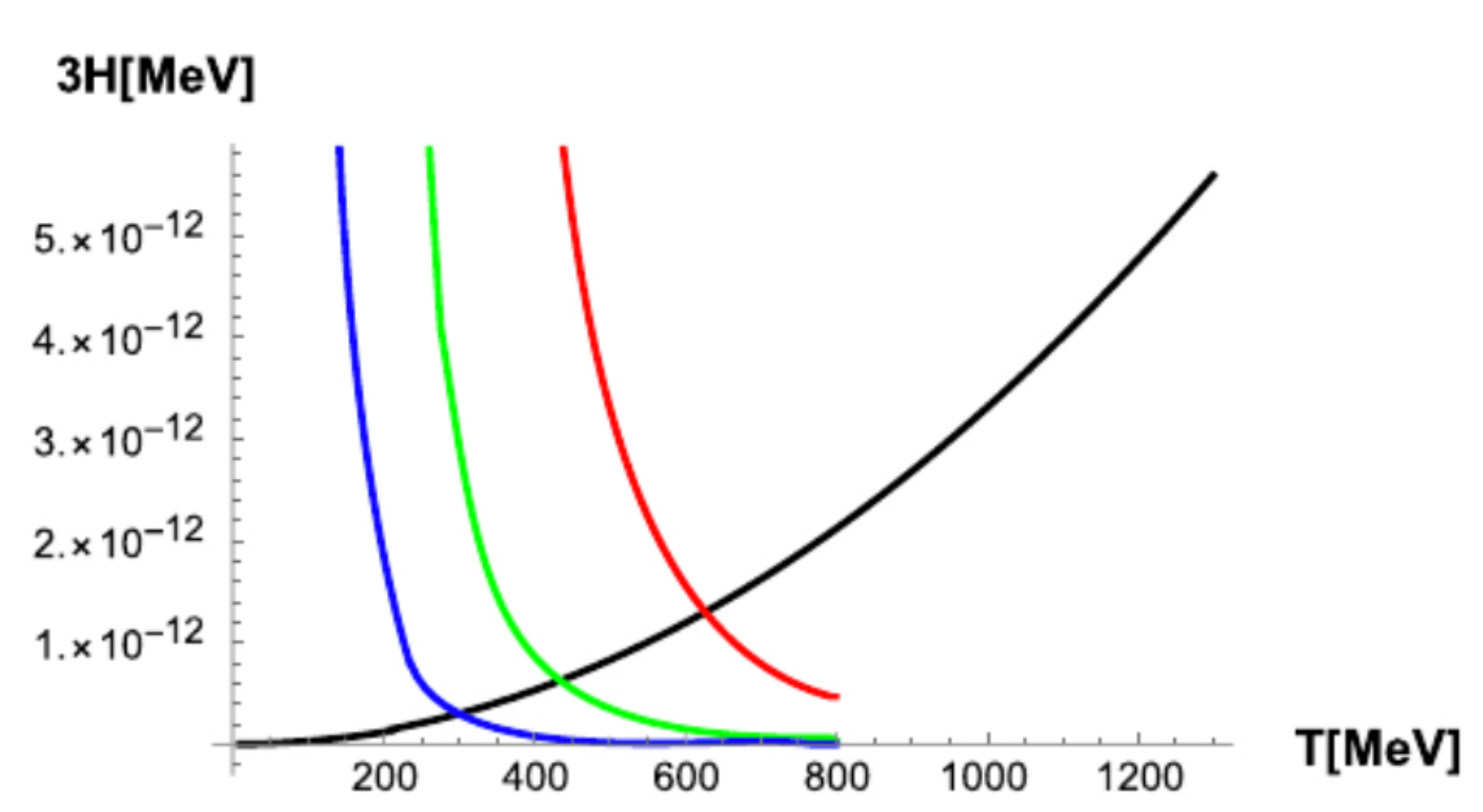}}
\caption{Determination of $T_1$ for $m_a(0)= 10^{-5}$eV(blue),  $10^{-4}$eV(green), and $ 10^{-3}$eV(red),  respectively.}\label{Fig:T1}
 \end{figure}
 %%%%%%%%%%%%%%%

Figure \ref{Fig:Overall} summarizes, by arrowed red curves, the evolution of  axion energy density  from the value  at $T_1$ until the finishing time $t_f$ of phase transition. One obvious boundary condition is that at the beginning of the phase transition it is 100\,\% in  the quark-gluon phase and we use the axion mass formula (\ref{eq:QgVar}) and 
pressure from (\ref{eq:AboveTc}). Without applying the evolution equation, these are used for estimating QCD variables. From $T_c$, go backward in time via blue curve to the high temperature regime. Different zero temperature axion masses have different blue curves.  
%%%%%%%%%%%%%%%%%%%
\begin{figure}[h!]
 {\includegraphics[width=0.7\textwidth]{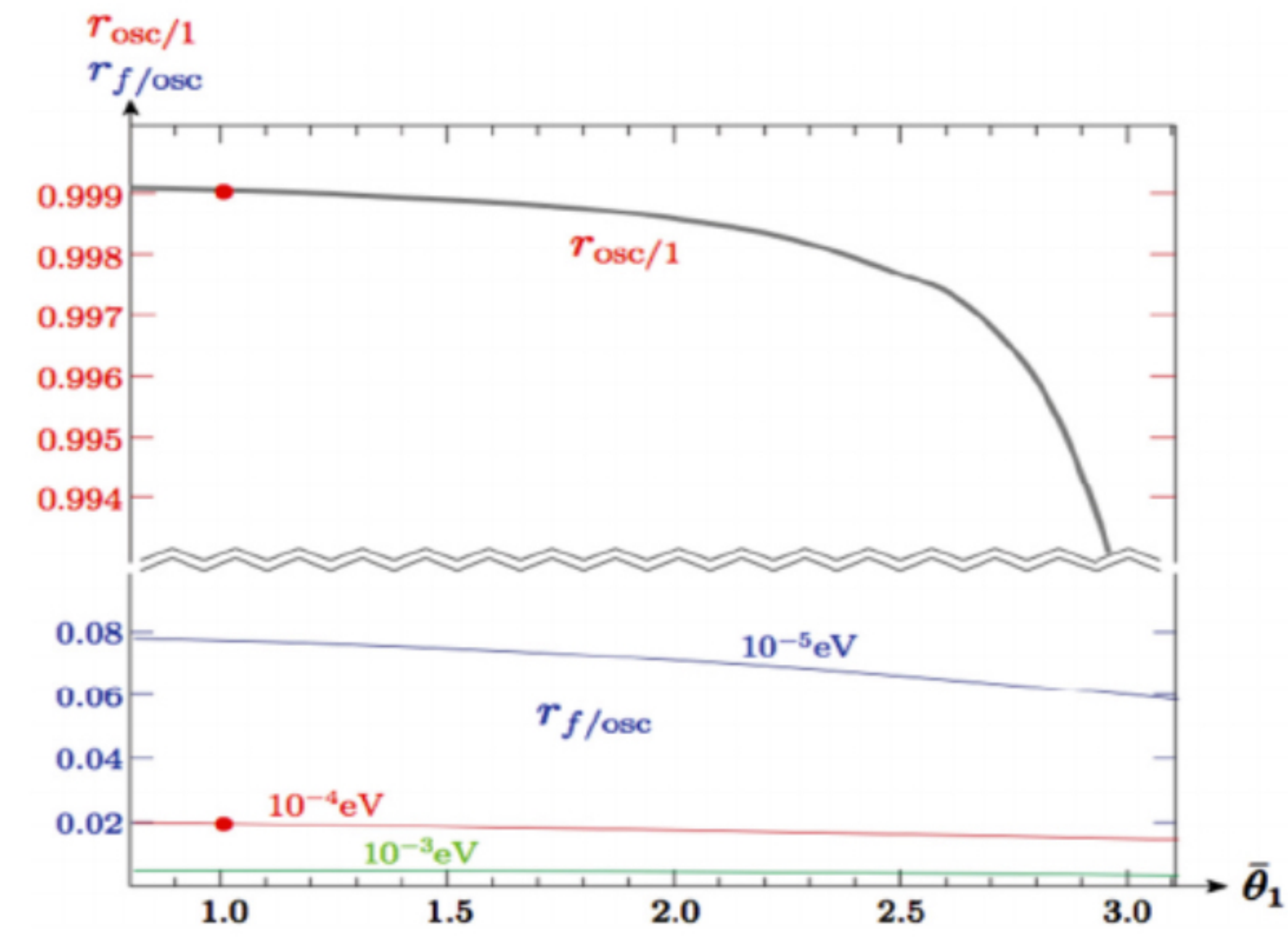}}
\caption{The ratio of $\bar{\theta}$'s for the initial $\bar{\theta}_1$. }\label{Fig:MassPower}
 \end{figure}
 %%%%%%%%%%%%%%%
These blue curve  masses in Fig. \ref{Fig:Overall} follow the afore-mentioned temperature powers, and in Fig. \ref{Fig:T1} we determine $T_1$ where $m_a(T_1)=3H(T_1)$.  We note that $T_1$ depends strongly on the zero temperature axion mass $m_a(0)$. For each $m_a(0)$ at the corresponding $T_1$, we calculate the axion energy density together with the evolution equation of $f_h$, shown as the solid red curve  in Fig. 4. The ratio of $\bar{\theta}$'s in terms of $\bar{\theta}_1$ is shown in Fig. 6, which is summarized as the axion mass dependence of the axion energy density or the current value of  $\bar{\theta}$ in terms of  $\bar{\theta}_1$ as
 \begin{equation}
 \bar{\theta}_{\rm now}=  r_{{\rm now}/f} \, \bar{\theta}_f,~~  \bar{\theta}_f\simeq 0.02 \left(\frac{m_a}{10^{-4\,}\rm eV} \right)^{-0.591\pm 0.008}\,\bar{\theta}_1
\end{equation}
where $r_{{\rm now}/f}$. The ratio of $\bar{\theta}$'s, now and at the closing time $t_f$ of the QCD phase transition, is calculated in \cite{Kim18ThNow}. In \cite{Kim18ThNow}, calculation has been achieved by using the reparametriation invariance of the differential equation $\frac{d^2}{dt^2}\bar{\theta}+3H \frac{d}{dt}\bar{\theta}+\frac{m_a^2}{2}\bar{\theta}\simeq 0$ where the harmonic oscillation has been effective, which is satisfied after $t_f$.
Collecting all these factors, we obtained
\begin{equation}
 \bar{\theta}_{\rm now}    \simeq 0.62 \times 10^{-18}\,\bar{\theta}_1.
\end{equation}

%%%%%%%%%%
\section{A GUT model from string compactification : Flipped SU(5)}
 
\begin{figure}[h!]
 {\includegraphics[width=0.55\textwidth]{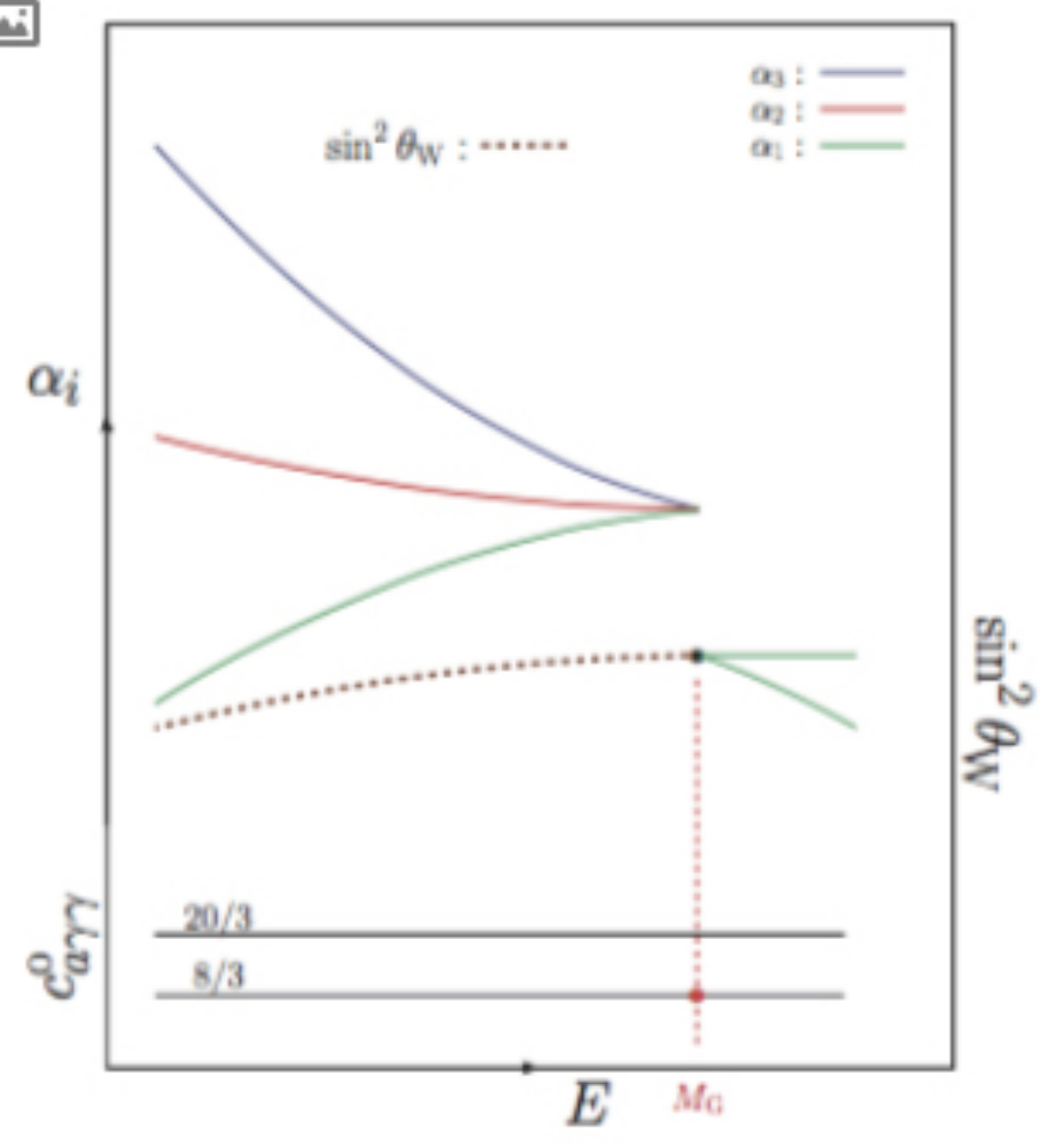}}
\caption{The scale dependences of gauge couplings and $\sin^2\theta_W$.  $c_{a\gamma\gamma}^0$ does not depend on the scale.   }\label{fig:GQW}
 \end{figure}

There is a useful  Georgi-Quinn-Weinberg(GQW) GUT relation  on the weak mixing angle $\theta_W$ in GUTs \cite{GQW74}: $\sin^2\theta_W={\rm Tr}T_3^2/{\rm Tr}Q_{\rm em}^2$. The GUT value $3/8$ was welcome in late 1970s with  $\sin^2\theta_W=0.233\pm 0.009\pm 0.005$ \cite{Kim81RMP}.\footnote{The most accurate current value is the LEP+SLD determination,  $0.23153\pm 0.00016 $ \cite{BodekA18}.}  An ``invisible'' axion-photon-photon coupling from GUT is usually the inverse of  $\sin^2\theta_W$, and $8/3$ is a standard GUT value for $c_{a\gamma\gamma}^0$ (without considering the QCD chiral symmetry breaking) which  below $T_c$ is shifted to $c_{a\gamma\gamma}\simeq c_{a\gamma\gamma}^0-2$. The scale dependences of  $\sin^2\theta_W$ and $c_{a\gamma\gamma}^0$ are shown in Fig. \ref{fig:GQW}.
 %%%%%%%%%%%%%%%%%%%
 %%%%%%%%%%%%%%%
 In this sense, the ``invisible'' axion is closely related to GUT models. In addition, the MI axion in string theory can survive as a global symmetry down to an intermediate scale   \cite{KimPatras17}.
 
The SU(5) GUT attracted a great deal of attention because of its simplicity. The branching of SO(10) can take the simplest route in the Georgi-Glashow (GG) model,
\begin{equation}
\textrm{SO(10)}\to \textrm{SU(5)}\times \textrm{U(1)}_X
\end{equation}
where all matter fields carry $X=0$ \cite{GG73}. The possibility of assigning nonzero $X$ to matter fields was noted by Barr, which has been named  `{\it flipped} SU(5)'  \cite{Barr82}, and its supersymmetric generalization without a need for an adjoint representation for GUT breaking was given in \cite{DKN84}. The $X$ charges in the GG  and  flipped SU(5) models are
\begin{eqnarray} 
\textrm{Georgi-Glashow}&:& X=0 ~\textrm{for all GUT representations},\\
\textrm{Flipped SU(5)}&:& {\rm Matter,~~}\overline{\bf 10}_{-1},~{\bf 5}_{+3},~{\bf 1}_{-5},\\
&& \textrm{Higgs,~~} {\bf 5}_{-2},~\overline{\bf 5}_{+2},\\
&& \textrm{Higgs,~~} {\bf 10}_{+1},~\overline{\bf 10}_{-1}.
\label{eq:Higgsflip}
\end{eqnarray} 
A good point of the flipped SU(5) from string is that  the level-1 fields ${\bf 10}_{+1}$ and $\overline{\bf 10}_{-1}$ needed for the GUT breaking are present in the compactification  \cite{flipSU5}. Even though, the standard-like model  \cite{KimJH07} from ${\bf Z}_{12-I}$, without the need for GUT breaking fields, is possible, this kind of standard-like models introduces too many chiral fields. So, to simplify the listing of all possible Yukawa couplings, it is desirable to have a flipped SU(5) from string.

%%%%%%%%%%
\section{Flavor symmetries}

 %%%%%%%%%%%%%%%%%
\begin{table}[t!]
\begin{center}
\begin{tabular}{@{}lc|cc|c|cccccc|c@{}} \hline
 &State($P+kV_0$)&$~\Theta_i~$ &${\bf R}_X$(Sect.)&$Q_R$ &$Q_1$&$Q_2$ &$Q_3$ &$Q_4$ &$Q_5$ &$Q_6$ & $Q_{\rm anom}$  \\[0.1em] \hline
 $\xi_3$  & $(\underline{+++--};--+)(0^8)'$&$0$ &$\tenb_{-1}(U_3)$&$+1$ &$-6$ & $-6$ & $+6$ & $0$ & $0$ & $0$ & $-13$   \\
$\bar{\eta}_3$  & $(\underline{+----};+--)(0^8)'$&$0$ &$\five_{+3}(U_3)$&$+1$  & $+6$ & $-6$ & $-6$ & $0$ & $0$ & $0$ & $-1$   \\
$\tau^c$  & $({+++++};-+-)(0^8)'$& $0$ &$\one_{-5}(U_3)$ &$+1$& $-6$ & $+6$ & $-6$ & $0$ & $0$ & $0$ & $+5$   \\
$\xi_2$  & $(\underline{+++--};-\frac{1}{6},-\frac{1}{6},-\frac{1}{6})(0^8)'$& $\frac{+1}{4}$ &$\tenb_{-1}(T_4^0)$& $-1$ &$-2$ & $-2$ & $-2$ & $0$ & $0$ & $0$ & $-3$ \\
$\bar{\eta}_2$  & $(\underline{+----};-\frac{1}{6},-\frac{1}{6},-\frac{1}{6})(0^8)'$&$\frac{+1}{4}$ &$\five_{+3}(T_4^0)$& $-1$   & $-2$ & $-2$ & $-2$ &$0$ & $0$ & $0$ & $-3$  \\
$\mu^c$  & $({+++++};-\frac{1}{6},-\frac{1}{6},-\frac{1}{6})(0^8)'$&$\frac{+1}{4}$ & $\one_{-5}(T_4^0)$& $-1$  & $-2$ & $-2$ & $-2$ &$0$ & $0$ & $0$ & $-3$  \\
$\xi_1$  & $(\underline{+++--};-\frac{1}{6},-\frac{1}{6},-\frac{1}{6})(0^8)'$&$\frac{+1}{4}$ &$\tenb_{-1}(T_4^0)$& $-1$  &$-2$ & $-2$ & $-2$ & $0$ & $0$ & $0$ & $-3$ \\
$\bar{\eta}_1$  & $(\underline{+----};-\frac{1}{6},-\frac{1}{6},-\frac{1}{6})(0^8)'$&$\frac{+1}{4}$ &$\five_{+3}(T_4^0)$& $-1$   &$-2$ & $-2$ & $-2$ & $0$ & $0$ & $0$ & $-3$ \\
$e^c$  & $({+++++};-\frac{1}{6},-\frac{1}{6},-\frac{1}{6})(0^8)'$&$\frac{+1}{4}$ & $\one_{-5}(T_4^0)$& $-1$  &$-2$ & $-2$ & $-2$ &$0$ & $0$ & $0$ & $-3$  \\[0.2em]
\hline
 $H_{uL}$  & $(\underline{+1\,0\,0\,0\,0};\,0\,0\,0)(0^5;\frac{-1}{2}\,\frac{+1}{2}\,0)'$&$\frac{+1}{3}$ &$2\cdot \five_{-2}(T_6)$& $-2$  & $0$ & $0$ & $0$ & $-12$ & $0$ & $0$ & $0$   \\
 $H_{dL}$  & $(\underline{-1\,0\,0\,0\,0};\,0\,0\,0)(0^5;\frac{+1}{2}\,\frac{-1}{2}\,0 )'$&$\frac{+1}{3}$ &$ 2\cdot \fiveb_{+2}(T_6)$&$+2$ &$0$ & $0$ & $0$ & $+12$& $0$& $0$& $0$   \\ [0.2em]
 \hline
$\Sigma^*_1$  & $(\underline{+++--};0^3)(0^5;\frac{-1}{4}\,\frac{-1}{4}\,\frac{+2}{4})'$ &$0$  & $2\,\tenb_{-1}(T_3)_L$ &$+4$& $0$ & $0$ & $0$ & $0$ & $+9$ & $+3$ & $\frac{-33}{7}$   \\
$\Sigma^*_1$  & $(\underline{+++--};0^3)(0^5;\frac{-1}{4}\,\frac{-1}{4}\,\frac{+2}{4})'$ &$\frac{+2}{3}$  & $1\,\tenb_{-1}(T_3)_L$ &$+4$& $0$ & $0$ & $0$ & $0$ & $+9$ & $+3$ & $\frac{-33}{7}$  \\
${\Sigma}_2$  & $(\underline{++---};0^3)(0^5;\frac{+1}{4}\,\frac{+1}{4}\,\frac{-2}{4})'$& $0$ &$2\, \ten_{+1}(T_3)_L$ &$-4$&$0$ & $0$ & $0$ & $0$ & $-9$ & $-3$ & $\frac{+33}{7}$   \\
 ${\Sigma}_2$  & $(\underline{++---};0^3)(0^5;\frac{+1}{4}\,\frac{+1}{4}\,\frac{-2}{4})'$& $\frac{+1}{3}$  &$1\, \ten_{+1}(T_3)_L$ &$-4$&$0$ & $0$ & $0$ & $0$ & $-9$ & $-3$ & $\frac{+33}{7}$  
 \\[0.2em]
\hline
 \end{tabular} \label{tab:SMfields} 
\end{center}
\caption{U(1) charges of matter fields in the flipped SU(5).  $\xi_i$ and $\bar{\eta}_i$ contain  the left-handed quark and lepton doublets, respectively, in the $i$-th family.}
\end{table}
%%%%%%%%%%%%%%

A flipped SU(5) with a small number of representations was discovered in \cite{Huh09}.  In Ref.  \cite{Kim18Fl1},  the Yukawa couplings were studied by assigning the third family members in the twisted sector $T_4^0$, which needed a fine-tuning to forbid the dimension-5 proton decay operators sufficiently. To resolve this problem, the discrete symmetry ${\bf Z}_{4R}$ has been found by assigning  the third family members in the untwisted sector $U$ \cite{Kim18Rp}.\footnote{This part is done after the talk at Corfu.} For reference, we list the non-singlet fields  in Table \ref{tab:SMfields}. It has been shown that the quark \cite{RpCKM} and lepton \cite{Kim19pmns} mixing angles from \cite{Kim18Rp} are possible to fit to the observed data.

In Table 1, $Q_R$ is the quantum number of discrete  ${\bf Z}_{4R}$ symmetry. Since  ${\bf Z}_{4R}$ is a subgroup of U(1)$_R$ symmetry, we need Yukawa coupling terms from F-terms satisfying $\sum_i Q_R^i=2$ modulo 4 and  Yukawa coupling terms from D-terms satisfying $\sum_i Q_R^i=0$ modulo 4. One can check that the fields in Table 1  alone cannot give any renormalizable Yukawa couplings. One needs more flipped SU(5) singlet fields $\sigma_i$ (listed in Refs. \cite{Kim18Rp, Kim19pmns}) to form non- renormalizable Yukawa couplings. In particular, all the needed Yukawa couplings are obtained by attaching the flipped SU(5) singlets to the well-known SM Yukawa couplings. Here, we just cite the mixing matrices and the CP violation magnitude $J$,
\begin{eqnarray}
V_{\rm CKM} &=&   \left(\begin{array}{ccc}
+0.974395 \cdot e  ^{i(8.90745\times 10^{-5}) }  ,+0.224814 \cdot e  ^{i(2.51923\times 10^{-5}) }  , +0.003615 \cdot e  ^{i(-\frac{\pi}{2} +0.4005) } \\[0.5em]
-0.224672 \cdot e  ^{i(6.302\times 10^{-4}) } ,+0.973517 \cdot e  ^{i(-7.666\times 10^{-5}) } ,+0.042275 \cdot e  ^{i(1.258\times 10^{-4}) }   \\[0.5em]
+0.008754 \cdot e  ^{i(-0.37945) } ,-0.041516 \cdot e  ^{i(1.788\times 10^{-2}) } ,0.99910 \cdot e  ^{i(-4.506\times 10^{-5}) } \\
 \end{array}\right),\nonumber\\[0.3em]
 &&J_{\rm CKM}=3.08\times 10^{-5}.\label{eq:fitCKM}
\end{eqnarray} 
and  
\begin{eqnarray}
V_{\rm PMNS} &=&  \left(\begin{array}{ccc} 0.82939,& 0.53909,&  0.14663 \\ [0.2em]
 -0.47985, & 0.68740+0.13441 e^{-i\delta} ,&
0.18697 -0.49417 e^{-i\delta}  \\[0.2em]
-0.28611 e^{i\delta} ,&-0.22543 +0.40986e^{i\delta}  ,& 0.82880 +0.11148 e^{-i\delta}
\end{array}\right),\nonumber\\[0.3em]
J_{\rm PMNS}&=&2.8838\times 10^{-2}\sin\delta\,.\label{eq:fitPMNS}
\end{eqnarray} 
In the quark sector, the CP violation has been determined rather accurately  and the phases are given in terms of numbers in Eq. (\ref{eq:fitCKM}). On the other hand, in the lepton sector the CP violation has not been determined yet  and the phases are given in terms of unknown $\delta$ in Eq. (\ref{eq:fitPMNS}).

%%%%%%%%%%
\section{Conclusion}
Key aspects of ``very light'' or  ``invisiblet'' axion working for a dark matter  possibility are discussed.  Most discussion was centered on the QCD phase transition from which the current value of $\overline{\theta}$ has been preseented. Our result is about 100 times larger than the previous quotations. The intermediate scale axion decay constant,  $f_a\sim 10^{11\,}$GeV, may be rooted in the string compactification. The model-independent axion in string theory with an anomalous U(1) gauge symmetry can be the origin of a global symmetry which is, by several applications of the 't Hooft mechanism at the GUT scale, broken at an intermediate scale. In this case, the dominant component of the  ``very light'' axion is housed in the SM singlet fields $\sigma_i$ from E$_8\times$E$_8'$, not in the second rank anti-symmetric tensor field $B_{MN}$.

Related to the GUT origin of the  ``very light'' axion, the flipped SU(5) from string is discussed. This  connects to the question on the flavors and we commented on the recent attempts to understand the quark and lepton mixing angles and the weak CP violation from the flipped SU(5).
 
%%%%%%%%%%%%%%%%% 


\begin{thebibliography}{99}

\def\apj#1#2#3{{Astrophys.\ J.}\ {\bf #1} (#3) #2}
\def\apjs#1#2#3{{Astrophys.\ J. Supp.}\ {\bf #1} (#3) #2}
\def\apph#1#2#3{{Astropart. Phys.}\ {\bf #1} (#3) #2}
\def\anj#1#2#3{{Astronom.\ J.}\ {\bf #1} (#3) #2}
\def\anap#1#2#3{{Astronom.\ Astrophys.}\ {\bf #1} (#3) #2}
\def\mnras#1#2#3{{Mon.\ Not.\ R.\ Astron.\ Soc.}\ {\bf #1} (#3) #2}
\def\anrnp#1#2#3{{Annu.\ Rev.\ Nucl.\ Part.\ Sci.}\ {\bf #1} (#3) #2}
\def\ijmpa#1#2#3{{Int.\ J.\ Mod.\ Phys. A}\ {\bf #1} (#3) #2}
\def\mpl#1#2#3{{Mod.\ Phys.\ Lett.\ A}\ {\bf #1} (#3) #2 }
\def\nat#1#2#3{{Nature}\ {\bf #1} (#3) #2}
\def\npb#1#2#3{{Nucl.\ Phys.\ B}\ {\bf #1} (#3) #2}
\def\plb#1#2#3{{Phys.\ Lett.\ B}\ {\bf #1} (#3) #2}
\def\prd#1#2#3{{Phys.\ Rev.\ D}\ {\bf #1} (#3) #2}
\def\prx#1#2#3{{Phys.\ Rev.\ X}\ {\bf #1} (#3) #2}
\def\pr#1#2#3{{Phys.\ Rev.}\ {\bf #1} (#3) #2}
\def\prl#1#2#3{{Phys.\ Rev.\ Lett.}\ {\bf #1} (#3) #2}
\def\prp#1#2#3{{Phys.\ Rep.}\ {\bf #1} (#3) #2}
\def\sjnp#1#2#3{{Sov.\ J.\ Nucl.\ Phys.}\ {\bf #1} (#3) #2}
\def\zp#1#2#3{{Z.\ Phys.}\ {\bf #1} (#3) #2}
\def\jhep#1#2#3{{JHEP}\ {\bf #1} (#3) #2}
\def\epjc#1#2#3{{Euro. Phys. J. C}\ {\bf #1} (#3) #2}
\def\rmp#1#2#3{{Rev. Mod. Phys.}\ {\bf #1} (#3) #2}
\def\sci#1#2#3{{Science}\ {\bf #1} (#3) #2}
\def\prth#1#2#3{{Prog. Theor. Phys.}\ {\bf #1} (#3) #2}
\def\njp#1#2#3{{New J. Phys.}\ {\bf #1} (#3) #2}
\def\mnra#1#2#3{{\it Mon.\,Not.\,Roy.\,Astron.\,Soc.} {\bf #1}, #2 (#3)}
\def\jkps#1#2#3{{J. Korean  Phys. Soc.}\ {\bf #1} (#3) #2}
\def\jetpl#1#2#3{{JETP Lett.}\ {\bf #1} (#3) #2}
\def\jcap#1#2#3{{JCAP}\ {\bf #1} (#3) #2}
  \def\cmp#1#2#3{{Comm. Math. Phys.}\ {\bf #1} (#3) #2}
\def\frp#1#2#3{{Front. Phys.}\ {\bf #1} (#3) #2}
\def\mpla#1#2#3{{Mod. Phys. Lett. A}\,{\bf #1} (#3) #2}
\def\err#1#2#3{Erratum: {\it ibid.} {\bf #1},  #2  (#3)}
 
%%%%%%%%%%%%%%%%%%%%%%%%%%%%%%%%%%%%%%%%%%%%%%%%%%%%%%%%%%%%%%%%%%

\bibitem{Mathur74}  J.E. Kim, V. S. Mathur, and S. Okubo, \emph{Electromagnetic properties of the neutrino from neutral-current experiments}, \prd{9}{3050}{1976} [doi:10.1103/PhysRevD.9.3050].

\bibitem{KimMM76}  J.E. Kim, \emph{Neutrino magnetic moment}, \prd{14}{3000}{1976} [doi:10.1103/PhysRevD.14.3000].

\bibitem{KimMM78}  J.E. Kim, \emph{Effects of the transition magnetic moment of the neutrino}, \prl{41}{360}{1978} [doi:10.1103/PhysRevLett.41.360].

\bibitem{Kim17Ch}  J.E. Kim, \emph{Naturally realized two dark Z's near the electroweak scale}, \prd{96}{055033}{2017} [arXiv:1703.10925 [hep-ph]].

\bibitem{Kim18IJMPA} J.  E. Kim, S. Nam, and Y. K. Semertzidis, \emph{Fate of global symmetries in the Universe:
QCD axion, quintessential axion and trans-Planckian inflaton decay constant}, \ijmpa{33}{1830002}{2017} [arXiv:1712.08648 [hep-ph]].

\bibitem{PQ77} R.D. Peccei and H.R. Quinn,  \emph{CP conservation in the presence of instantons},  \prl{38}{1440}{1977} [doi: 10.1103/PhysRevLett.38.1440].

\bibitem{Weinberg77} S. Weinberg, \emph{Conference Summary}, in Ben Lee Memorial Int. Conf. on Parity Nonconservation, Weak Neutral Currents and Gauge Theories, 20-22 Oct. 1977, ed. D. Cline and F. E. Mills (Harwood Academic Pub., London, 1978) p.727, and  
 F. Wilczek, \emph{Some problems in Gauge Field Theories}, {\it ibid} p.607.
 
 \bibitem{Peccei78} R. D. Peccei, \emph{A short review of axions}, in Proc. 19th ICHEP, eds. S. Homma, M.
Kawaguchi and H. Miyazawa, Tokyo, Japan, 23-30 August 1978 (Phys. Soc. of Japan,
Tokyo, 1979), p. 385.

\bibitem{Kim79}  J.E. Kim, \emph{Weak interaction singlet and strong CP invariance}, \prl{43}{103}{1979} [doi: 10.1103/PhysRevLett.43.103].

\bibitem{Kim87prp} J. E. Kim,  \emph{Light pseudoscalars, particle physics and cosmology}, \prp{150}{1}{1987} [doi: 
 10.1016/0370-1573(87)90017-2].

\bibitem{SVZ80} M. A. Shifman, A. I. Vainshtein and V. I. Zakharov, \emph{Can confinement ensure natural CP invariance of strong interactions?}, \npb{166}{4933}{1980}
[doi:10.1007/s10714-008-0707-4].

\bibitem{DFSZ} M. Dine, W. Fischler and M. Srednicki, \emph{Simple solution to the strong CP problem with a harmless axion}, \plb{104}{199}{1981} [ doi:10.1016/0370-2693(81)90590-6];  A. P. Zhitnitsky, \emph{On possible suppression of the axion hadron interactions} (in Russian), Yad. Fiz. {\bf 31} (1980) 497 [Sov. J. Nucl. Phys. {\bf 31} (1980) 260].

\bibitem{KimKyaeNam17} J. E. Kim, B. Kyae, and S. Nam, \emph{The anomalous U(1)$_{\rm anom}$ global symmetry and flavors from an SU(5)$\times$SU(5)$'$
GUT in ${\bf Z}_{\rm 12-I}$ orbifold compactification}, \epjc{77}{847}{2017}  [arXiv: 1703.05345 [hep-ph]]. 

\bibitem{Hooft71} G. 't Hooft,  \emph{Renormalizable Lagrangians for massive Yang-Mills fields}, \npb{35}{167}{19871} [doi:10.1016/0550-3213(71)90139-8].

\bibitem{KimPatras17} J. E. Kim, \emph{'t Hooft mechanism, anomalous gauge U(1), and ``invisible" axion from string}, Talk presented at Patras-17 [arXiv:1710.08454 [hep-ph]].

\bibitem{Kim88} J.E. Kim, \emph{The strong {CP} problem in orbifold compactifications and an SU(3)$\times$SU(2)$\times$U(1)$^n$}, \plb{207}{434}{1988} [doi:10.1016/0370-2693(88)90678-8].
 
\bibitem{GS84} M. Green and J. Schwarz, \emph{Anomaly cancellation in supersymmetric D=10 gauge theory and superstring theory}, \plb{149}{117}{1984} [doi:10.1016/0370-2693(84)91565-X].

\bibitem{Witten84} E. Witten, \emph{Some properties of O(32) superstrings }, \plb{149}{351}{1984} [doi:
 10.1016/ 0370-2693(84)90422-2].

 \bibitem{AnomU187} M. Dine,  N. Seiberg, and E. Witten, \emph{Fayet-Iliopoulos terms in string theory}, \npb{289}{589}{1987} [doi: 10.1016/0550-3213(87)90395-6]; 
J. J. Atick, L. Dixon, and A. Sen, \emph{String calculation of Fayet-Iliopoulos d terms in arbitrary supersymmetric compactifications}, \npb{292}{109}{1987} [doi:10.1016/0550- 3213(87)90639-0];
M. Dine, I. Ichinose, and N. Seiberg, \emph{F terms and d terms in string theory}, \npb{293}{253}{1987} [doi:10.1016/0550-3213(87)90072-1].

\bibitem{ICTP16} G. Grilli di Cortona, E. Hardy, J. P. Vega, and G. Villadoro, \emph{The QCD axion, precisely}, \jhep{1601}{034}{2016} [arXiv:1511.02867 [hep-ph]].

\bibitem{Borsanyi16} Sz. Borsanyi {\it et al.}, \emph{Calculation of the axion mass based on high-temperature lattice quantum chromodynamics}, \nat{539}{69}{2016}  [arXiv:1606.07494 [hep-lat]].

\bibitem{Preskill83} J. Preskill, M. B. Wise, and F. Wilczek, \emph{Cosmology of the invisible axion}, \plb{120}{127}{1983} [doi: 10.1016/0370-2693(83)90637-8];
  L. F. Abbott and  P. Sikivie, \emph{A cosmological bound on the invisible axion}, \plb{120}{133}{1983} [doi:10.1016/0370-2693(83)90638-X]; M. Dine and W. Fischler, \emph{The not so harmless axion},  \plb{120}{137}{1983} [doi:10.1016/0370-2693(83)90639 -1]. 

\bibitem{DeGrand84} T. DeGrand and K. Kajantie, \emph{Supercooling, entropy production, and bubble kinetics in the  
quark-hadron phase transition in the early universe}, \plb{147}{273}{1984} [doi: 10.1016/ 0370-2693(84)90115-1].

\bibitem{KolbBk} E. W. Kolb and M. S. Turner, \emph{The Early Universe} (Addison-Wesley Pub. Co., New York, 1990).
 
\bibitem{HuangBk} K. Huang, \emph{Introduction to Statistical Physics} (Taylor \& Francis, London, 2001).

\bibitem{Kim18QCD} J. E. Kim and  S-J. Kim,  \emph{``Invisible'' QCD axion rolling through the QCD phase transition}, \plb{783}{357}{2018} [arXiv:1804.05173 [hep-ph]].

\bibitem{Baluni79} V. Baluni, \emph{CP violating effects in QCD}, \prd{19}{2227}{1979} [doi:10.1103/ PhysRevD. 19.2227].

\bibitem{Pisarski81} D. J. Gross, R. D. Pisarski, and L. G. Yaffe, \emph{QCD and instantons at finite temperature}, \rmp{53}{43}{1981} [doi:10.1103/RevModPhys.53.43].

\bibitem{Kim18ThNow}  J. E. Kim, S-J. Kim, and S. Nam, \emph{Axion energy density, bottle neck period, and $\bar{\theta}$ ratios between early and late times}, 
[arXiv:1803.03517 [hep-ph]].

\bibitem{GQW74} H. Georgi, H. R. Quinn, and S. Weinberg, \emph{Hierarchy of interactions in unified gauge theories}, \prl{33}{451}{1974} [doi:10.1103/PhysRevLett.33.451].

\bibitem{Kim81RMP} J. E. Kim, P. Langacker, M. Levine, and H. H. Williams, \emph{A theoretical and experimental review of the weak neutral current: A determination of its structure and limits on deviations from the minimal SU(2)$_L \times $U(1) electroweak theory}, \rmp{53}{211}{1981} [doi:10.1103/RevModPhys. 53.211].

\bibitem{BodekA18} For a recent review, see, A. Bodek,  \emph{Measurement of effective weak mixing angle $\sin^2\theta_{eff}^{lept}$ from the forwrd-backward asymmetry of Drell-Yan events at CMS}, Talk presented at CIPANP2018, the 13th International conference on the intersections of particle and nuclear physics, Palm Springs, California May 28--June 3, 2018, [arXiv:1808.03170].  

\bibitem{GG73} H. Georgi and S. L. Glashow, \emph{Unity of all elementary particle forces}, \prl{32}{438}{1974} [doi:10.1103/PhysRevLett.32.438].
 
 \bibitem{Barr82} S. M. Barr, \emph{A new symmetry breaking pattern for SO(10) and proton decay}, \plb{112}{219}{1982}  [doi: 10.1016/0370-2693(82)90966-2].
 
 \bibitem{DKN84} J. P. Derendinger, J.  E. Kim, and D. V. Nanopoulos, \emph{Anti-SU(5)}, \plb{139}{170}{1984} [doi: 10.1016/0370-2693(84)91238-3].

\bibitem{flipSU5} I.  Antoniadis, J.\,R. Ellis, J\,.S. Hagelin, and D.\,V. Nanopoulos, \emph{The flipped SU(5)$\times$U(1) string model revamped}, \plb{231}{65}{1989} [doi: 10.1016/0370-2693(89)90115-9]; J. E. Kim and B. Kyae, \emph{Flipped SU(5) from ${\bf Z}_{12-I}$ orbifold with Wilson line}, \npb{770}{47}{2007} [arXiv: hep-th/ 0608086].

   
\bibitem{KimJH07} J. E. Kim, J-H. Kim, and B. Kyae,  \emph{Superstring standard model from ${\bf Z}_{12-I}$ orbifold compactification with and without exotics, and effective R-parity}, \jhep{0706}{034}{2007}  [arXiv: hep-ph/0702278].

\bibitem{Huh09}  J-H. Huh, J. E. Kim, and B. Kyae, \emph{SU(5)$_{\rm flip}\times SU(5)^\prime$ from  ${\bf Z}_{12-I}$},  \prd{80}{115012}{2009} [arXiv:0904.1108 [hep-ph]]. 

\bibitem{Kim18Fl1} J. E. Kim, \emph{Theory of flavors: String compactification}, \prd{98}{055005}{2018} [arXiv: 
 1805.08153[hep-ph]]. 
     
\bibitem{Kim18Rp} J. E. Kim, \emph{R-parity from string compactification}, arXiv:1810.10796.

\bibitem{RpCKM} J. Jeong, J. E. Kim, and S-J. Kim, \emph{Flavor mixing inspired by flipped SU(5) GUT}, arXiv:1812.02556 [hep-ph].

\bibitem{Kim19pmns} J. Jeong, J. E. Kim, and S. Nam, \emph{Leptonic CP violation in flipped SU(5) GUT from ${\bf Z}_{12-I}$ orbifold compactification}, arXiv:1901.02295 [hep-ph].
 
\end{thebibliography}
\end{document}